\documentclass[a4paper,12pt]{article}
\usepackage{latexsym}
\usepackage{amsmath,amssymb,bm,ascmac,color,calc}
\usepackage[mathscr]{eucal}
\usepackage{graphicx}
\usepackage{cite}
\usepackage{lineno}
\usepackage{setspace}
\usepackage{etoolbox}
\usepackage{url}


\allowdisplaybreaks 
\usepackage{longtable} 

\begin{document}
\begin{titlepage}

\vspace*{2cm}

\begin{center}
{\large\bf 
Implementation of muon pair production in PHITS and verification by comparing with the muon shielding experiment at SLAC
}\\[15mm]


\renewcommand{\thefootnote}{\fnsymbol{footnote}}
Yasuhito Sakaki, Yoshihito Namito, Toshiya Sanami, Hiroshi Iwase, Hideo Hirayama  \\ \bigskip
\renewcommand{\thefootnote}{\arabic{footnote}}

{\em KEK Radiation Science Center, Tsukuba, Ibaraki 305-0801, Japan}
\end{center}

\vspace{1cm}
\begin{abstract}
We implemented a model of muon pair production through a real photon in {\tt PHITS} and compared our calculations with data of the muon shielding experiment conducted at SLAC to verify the validity of the implemented model.  Our predictions of the muon fluence induced by electrons are in good agreement with the experimental data. To understand the known differences between the calculations of the muon fluence, which have been determined using other Monte-Carlo codes, we quantitatively evaluate the fluctuations in the Monte-Carlo results due to systematic errors in multiple Coulomb scattering, differences in the approximation methods and energy loss models, and whether incoherent production is considered.

\end{abstract}
\end{titlepage}


\setlength{\parskip}{10pt}%

\section{Introduction}

Accurate particle transport calculations are essential for realizing adequate designs of equipment and facilities that employ accelerated beams. For this purpose, integrated Monte-Carlo codes are currently widely used.  
In terms of radiation shielding, a high precision calculation tool is necessary in order to avoid creation of a redundant design of the shielding used to mitigate radioactive contamination, and radiation exposure. 
Currently, the energy and intensity of beams used in accelerated facilities are increasing to obtain scientifically significant data, and the worth of reliable Monte-Carlo calculations over a wide energy range is rising.

In an accelerator facility, when a photon, electron, or positron of about 210 MeV or more is injected into matter, it interacts with a nucleus in the matter and generates a muon pair, which consists of a negatively and positively charged muon.  These muons are emitted in the forward direction with respect to the direction of the incident particles, and they form high-dose sections over long distances owing to their high penetrating power.  This property of muons often causes difficulties in facility design, and one of the problems associated with inadequate design is muon-induced radioactivity.

Muon activation is mainly caused by negative muon capture and electromagnetic interactions between muons and nuclei.  This issue is of particular concern around beam dumps and collimators in future accelerators using high-intensity and high-energy electron-positron beams.  
In the case of electron-positron linear accelerators, high-intensity and high-energy muon fields can be formed behind beam dumps, as most accelerated particles are injected into the beam dumps.  The muon produces long-lived radionuclides when it passes through objects like concrete, rock and groundwater. As a consequence, we have to consider muon shielding as necessary.

In addition to the radio-activation, muons may contribute to a background in experiments.  In electron-positron colliders, accelerated particles hit a collimator in the upstream section of a collision point, and muons generated in this section can reach the detectors and become background in the results of experimental measurements~\cite{Drozhdin:2007zz}.  As another example, pair-produced muons are the main parts of the background in dark matter search experiments utilizing electron beam dumps, such as ones been planning at the Jefferson Lab and SLAC~\cite{Raubenheimer:2018mwt,Battaglieri:2019nok}.  
To deal with these kinds of phenomena, it is necessary to predict the absolute value and angular spread of the muon field with a high accuracy.

This study serves two purposes.  The first one is to implement a muon pair production model in a Monte-Carlo code {\tt PHITS}~\cite{sato:PHITS}.  The current version of {\tt PHITS}\footnote{{\tt PHITS-3.17} } does not include the production model.  This implementation will be released in the near future.  We employ a differential cross section given by a perturbative calculation of quantum electrodynamics and consider three physical modes for the process.  
We, herein, report the details of the muon generation model implemented in {\tt PHITS}.  
The second purpose is to verify the results of the updated code with data of the muon shielding experiment at SLAC~\cite{Nelson:1974tu, Nelson:1974tv}.  This experiment is very useful for verifying Monte-Carlo calculations on the muon field originating from high energy electrons, because the experimental geometry is clear.  Also, data on the absolute value and angular spread of the muon flux are shown, which are significant observables in terms of muon shielding.  This data has been used as a benchmark for Monte-Carlo calculations on muon fields of electron origin.  Although a previous study has confirmed a difference of more than 10 times between the results of several Monte-Carlo codes in the large angle region, we do not yet know the cause for this~\cite{Fasso:2017cul}.  In this paper, we consider muon generation and transport in detail and hope that our results are a clue for understanding the causes of the differences between other Monte-Carlo codes.

This paper is organized as follows.  
In Sec.~\ref{sec:Muon_pair_production}, we summarize theoretical aspects of the muon pair production and the differential cross-section implemented in the code.  
In Sec.~\ref{sec:Results}, we show numerical results and compare them with data on the SLAC muon shielding experiment.  The changes or uncertainties on muon fluence owing to some changes in calculation options and are quantitatively evaluated. 
In Sec.~\ref{sec:Summary}, we summarize our results and state our conclusions.

\section{Muon pair production}
\label{sec:Muon_pair_production}

In this section, we summarize the theoretical aspects of the muon pair production and the differential cross section implemented in {\tt PHITS}.  This first implementation contains the major contributions to muon pair production.  Minor contributions, which are not included in the code are detailed in this section too.  Users of {\tt PHITS} can include these contributions by modifying a source file themselves.  If there are any official updates, they will be described in Sec.~1 of the {\tt PHITS} manual~\cite{ref:phits}.

In the muon pair production process, an incident photon produces a negatively and positively charged muon pair mainly through electromagnetic interaction with the nucleus, nucleons and partons.  This process is categorized into two types, depending on whether the incident photon is a real photon or a virtual photon.  In this work, we consider only the production originating from a real photon, which is usually the dominant contribution.  When a high energy electron beam is injected into matter, it generates a lot of real photons (bremsstrahlung photons) in an electromagnetic shower, and then, the photons generate muon pairs.  This is the main muon production process of matter with electron beams.  In contrast, an electron can interact directly with a nucleus or nucleon, in which a virtual photon (mediator of the electromagnetic force) produces a muon pair.
We ignore the contribution from the virtual photon as this contribution is much smaller than that of the real photon when the thickness of the medium into which the electron is injected is more than 1/50 times of the radiation length~\cite{Tsai:1973py}.

\begin{figure}[!ht]
\begin{center}
\includegraphics[width=7.0cm, bb=0 0 790 352]{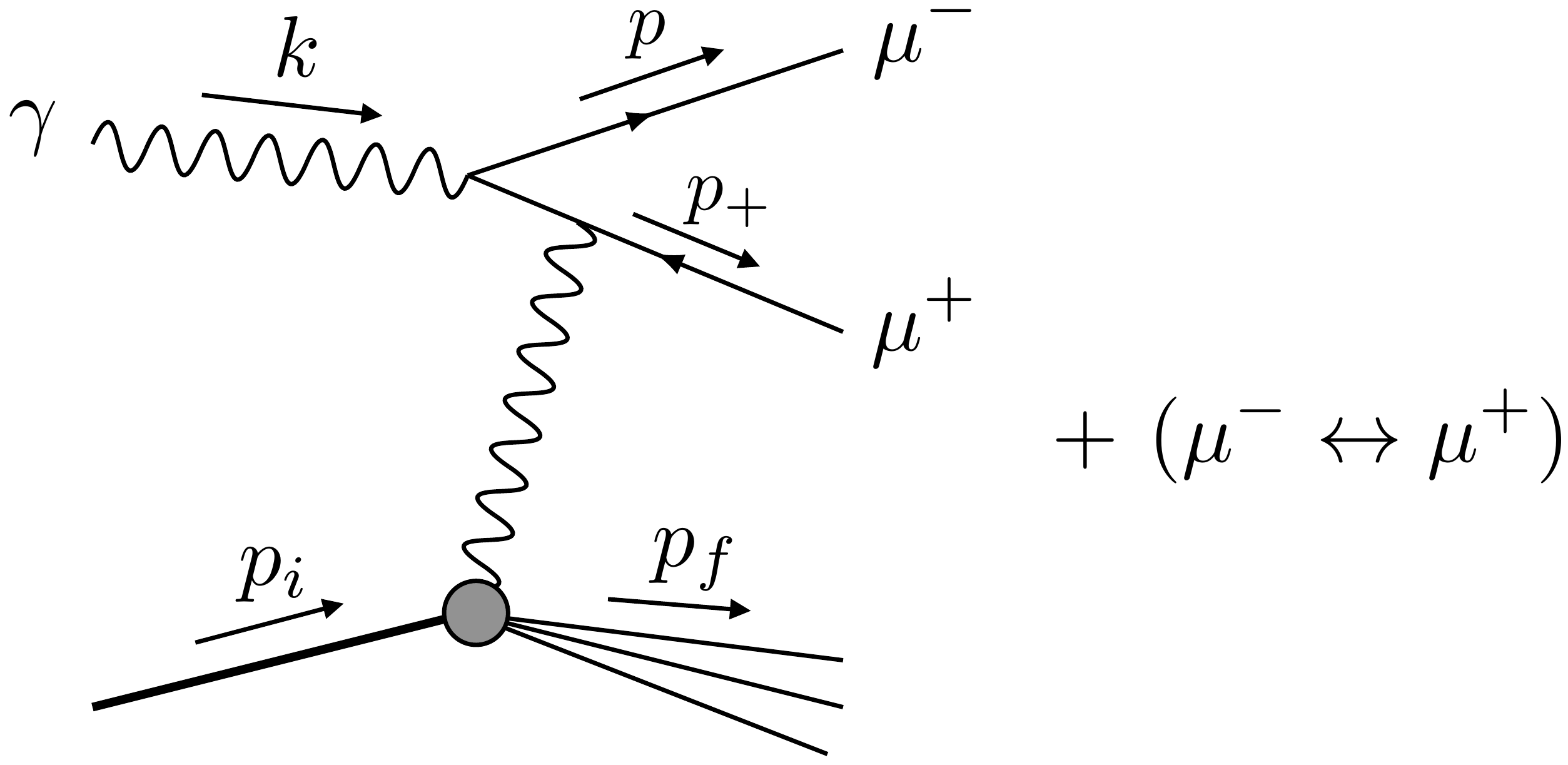}
\caption{{\footnotesize A Feynman diagram for the Bethe-Heitler process.}}
\label{BHdiagram}
\end{center}
\end{figure}
We consider only the Bethe-Heitler process in Figure~\ref{BHdiagram}, which has a greater contribution than other real photon processes.  In this process, a muon interacts electromagnetically with a nucleus or nucleon.  In the perturbation theory of quantum electrodynamics, we basically expand the scattering amplitude in a power series of the fine structure constant $\alpha$, and this diagram gives the largest contribution at the order of $\alpha^3$.  It is known that the virtual Compton scattering process has a contribution at the same order~\cite{Krass:1965hh}, but we neglect this process.  This is because the Bethe-Heitler process has a singularity of $1/t$, where $t$ is the momentum transfer (see definition below), and also has a collinear singularity for the muon angle in the  high energy limit, then the Bethe-Heitler process is more significant than the virtual Compton scattering.

We use a formula of the differential cross section for the Bethe-Heitler process in the Born approximation based on Ref~\cite{Tsai:1973py}, which is based on a perturbative calculation in quantum electrodynamics.  In the Bethe-Heitler process, we consider three physical modes ($\mathcal{P} \in$ \{coherent production, quasi-elastic scattering, inelastic scattering\}), and then, the total cross section is given by the sum of these three modes ($\Sigma_{\mathcal{P}}$) as
\begin{align}
\frac{d\sigma}{dp d\Omega} 
&= -\frac{\alpha^3}{2 \pi}  \sum_{\mathcal{P}}
  \int _{t_{\min}}^{t_{\max}}\frac{dt}{2 p_{is} p_{+s}}
  \int _{m_f^2}^{(U-m)^2} dm_f^2 \frac{p^2}{U k E} \frac{p_{+s}}{t^2} \label{eq:Born} \\
&\times \left[ 
         W_2^{(\mathcal{P})}(t,m_f^2) \left(\frac{H W}{Y^3 k_s} +\frac{B}{Y k_s} +C +D k_s W \right)
              \right. \nonumber\\
&+         \left. 
         W_1^{(\mathcal{P})}(t,m_f^2) \left(\frac{H' W}{Y^3 k_s} +\frac{B'}{Y k_s} +C' +D' k_s W \right)
              \right],  \nonumber
\end{align}
where, 
\begin{itemize}
\setlength{\parskip}{0cm}
  \item[] $k$ is the four-momentum or energy of the incident photon in the laboratory frame,  
  \item[] $p$ $(p_+)$ is the four-momentum or absolute value of the three-momentum in the laboratory frame for a negative (positive) muon, 
  \item[] $p_i$ is the four-momentum of a target particle, 
  \item[] $p_f$ is the four-momentum sum of all final states except for muons. 
\end{itemize}
Therefore, four-momentum conservation is given by $k+p_i=p+p_+ +p_f$.  Moreover, 
\begin{align}
&t =-(k-p-p_+)^2\\
& ~ =-2 m_{\mu}^2  +2 k\cdot p +2E_{+s} (k_s-E_s) -2p_{+s}p_{is}\cos\theta_{+}, \label{eq:t}\\
&m_f^2=p_f^2, \\
&E_{(+)}=\sqrt{p_{(+)}^2+m_{\mu}^2}, \\
&E_{+s}=\sqrt{p_{+s}^2+m_{\mu}^2} 
             =\frac{U}{2}\left( 1+\frac{m_{\mu}^2-m_{f}^2}{U^2} \right), \label{eq:E+s} \\
&p_{is}=\frac{m_i}{U}\sqrt{k^2+p^2-2pk\cos\theta}, \label{eq:pis} \\
&k_s=\frac{k m_i -k\cdot p}{U}, \label{eq:ks} \\
&H=m_{\mu}^2 \left[ \frac{t}{2} \left(1-\frac{2E}{m_i}\right) -2E^2-2E \Delta \right], \\
&B= 
-\frac{2}{k\cdot p}\left\{ \left(m_{\mu}^2+\frac{t}{2} \right)
   \left[2E(E-k)-\frac{t}{2}\left( \frac{k-2E}{m_i}+1 \right)+(2E-k)\Delta\right]
   +\frac{t k^2}{2} \right\} \nonumber \\
   &\hspace{22pt}
    -\frac{t}{m_i}\left(m_i+E-k+\frac{t}{2m_i}\right) -2\Delta\left(\Delta-k+E+\frac{t}{m_i}\right) +k\cdot p , \\
&C=-\frac{m_{\mu}^2}{(k\cdot p)^2} \left[ 2\left(k-E-\Delta-\frac{t}{2}\right)(k-E)-\frac{t}{2} \right] 
      +\frac{1}{k\cdot p}\left[ 2E\Delta-t\left(1-\frac{E}{m_i}\right) \right], \\
&D=\frac{1}{k\cdot p}, \\
&H'=m_{\mu}^2(2m_{\mu}^2-t), \\
&B'=-\frac{t^2-4m^4}{k\cdot p}+2t-2k\cdot p-4m_{\mu}^2, \\
&C'=\frac{(m_{\mu}^2-2k\cdot p)(2m^2-t)}{(k\cdot p)^2}, \\
&D'=-\frac{2}{k\cdot p}, \\
&W=E_{+s}-p_{+s}\cos\theta_+\theta_k, \\
&Y=\sqrt{m^2\sin^2\theta_k+(p_{+s}\cos\theta_+ - E_{+s}\cos\theta_k)^2}, \\
&\cos\theta_k=(k_s-E_s)/p_{is}+(k\cdot p)/(k_s p_{is}), \\
&\Delta=\frac{m_f^2-m_i^2}{2m_i}, \\
&U^2=(p_+ + p_f)^2=m_{\mu}^2+m_i^2+2m_i(k-E)-2k\cdot p, 
\end{align}
where $m_{\mu}$ is the muon mass, $m_i$ is the target mass, $m_f$ is the invariant mass of the final state except muons, the subscript $s$ in Eqs.~(\ref{eq:E+s})-(\ref{eq:ks}) refers to the rest frame of $U$, and $t_{\rm max}$ and $t_{\rm min}$ are given by the settings $\cos\theta_+=-1$ and $1$ in Eq.~(\ref{eq:t}) respectively.  The target mass depends on the mode considered.

In coherent production, the incident photon interacts with the nucleus, then, the target mass is given by its mass, namely $m_i$ = (mass of target nucleus). The form factors for coherent production are given by 
\begin{align}
& W_1^{\text{(coherent)}}=0, \\
& W_2^{\text{(coherent)}}= 2m_i \delta(m_f^2 - m_i^2) Z^2 / (1+t/d)^2, \label{W2_coh}
\end{align}
where $d = 0.164A^{-2/3} {\rm GeV}^2$, and $Z$ and $A$ are the atomic number and mass number of the target.  What we should emphasize is the dependency of the momentum transfer $t$ in the form factor.  This variable corresponds to an {\it energy scale} of the elementary process, and $1/\sqrt{t}$ is proportional to an effective wavelength of a force mediator which couples with the nucleus.  As $t$ increases, the force mediator couples with a partial electric charge rather than the entire electric charge of the target nucleus, thus, the cross section for coherent production decreases.  In other words, coherent production prefers small-$t$ phase space.  As the muon angle increases, the small-$t$ phase space decreases, and thus, the muon emission angle tends to be smaller for coherent production than in other modes.  We will see this behavior in Figure~\ref{fluence_others_ratio}.

In the other two incoherent productions (quasi-elastic scattering, inelastic scattering), 
the incident photon interacts with the nucleon, rather than the nucleus. Their form factors are given by
\begin{align}
&W_j^{\text{(quasi-elastic)}} = C(t) \left[ Z W_{j p}^{\text{(el)}} + (A-Z) W_{j n}^{\text{(el)}}  \right], 
\quad j \in \{1,2\}, \\
&W_j^{\text{(inelastic)}} = Z W_{j p}^{\text{(meson)}} + (A-Z) W_{j n}^{\text{(meson)}}, 
\quad j \in \{1,2\}, \\
&C(t)=\begin{cases}
  \frac{3Q}{4P_F}\left( 1-\frac{Q^2}{12P_F^2} \right) & \text{if } Q<2P_F \\
  1 & \text{otherwise} \\
\end{cases},
\end{align}
where $Q^2=t^2/(2m_p)^2+t$, $P_F=0.25~{\rm GeV}$, and $m_p$ is the proton mass.  
The elastic and meson production form factors for the proton and neutron are given by
\begin{align}
& \left(
    \begin{array}{c}
      W_{2p}^{\text{(el)}}  \\[5pt]
      W_{1p}^{\text{(el)}}  \\[5pt]
      W_{2n}^{\text{(el)}}  \\[5pt]
      W_{1n}^{\text{(el)}}  
    \end{array}
  \right)
  = \frac{2m_p\delta(m_f^2-m_i^2)}{(1+t/0.71{\rm GeV}^2)^4}
\left(
    \begin{array}{l}
      (1+2.79^2\tau)/(1+\tau)  \\[5pt]
      2.79^2\tau  \\[5pt]
      1.91^2\tau/(1+\tau)  \\[5pt]
      1.91^2\tau
    \end{array}
  \right), \\
& W_{1p}^{\text{(meson)}}  = 
     C_M\left[ \frac{m_{\rho}^4(m_f^2-m_p^2)}{(m_{\rho}^2+t)^2}\frac{0.251}{{\rm GeV}^2}
     +\frac{m_p^2(1-x)^4}{1-1.26x+0.96x^2}\frac{0.646}{{\rm GeV}^2} \right] \\
& W_{1n}^{\text{(meson)}} \simeq W_{1p}^{\text{(meson)}}, \\
& W_{2p}^{\text{(meson)}}  = 
      \frac{1}{1+\frac{\nu^2}{t}} 
      \left[W_{1p} 
          +C_M \left(1-\frac{t}{2m_p\nu}\right)^2 
          \frac{(m_f^2-m_p^2)tm_{\rho}^2}{(m_{\rho}^2+t)^2}\frac{0.145}{{\rm GeV}^2}
       \right] \\
& W_{2n}^{\text{(meson)}} \simeq W_{2p}^{\text{(meson)}}, 
\end{align}
where $C_M=1/(8\pi^2 \alpha m_p)$, $\nu=(m_f^2-m_p^2+t)/(2m_p)$, $x=(1+m_f^2/t)^{-1}$, $\tau=t/(4m_p^2)$, and $m_{\rho}$ is the rho meson mass.  In our numerical calculations for these modes, the target mass is simply set to the proton mass, namely $ m_i = m_p$.

There are two things to notice about the inelastic form factors.  First, although inelastic form factors actually oscillate with changes in $m_f^2$ due to the existence of meson resonances, they are fitted using an average value.  We are interested in muon kinematics and observables integrated over $m_f^2$, thus, the above form factors work well for our purposes.  Second, it is necessary to consider the quark's and gluon's parton distribution functions in the region of deep inelastic scattering $(t>1{\rm GeV}^2)$.  We neglect the contributions of the inelastic form factors, because the cross section in this region is much smaller than the total cross section.  If users of PHITS want to know a realistic $m_f^2$ distribution and the cross section in the region of the deep inelastic scattering, this becomes possible by modifying the formulas of the inelastic form factors in a source file.

The nuclear (coherent) form factor in Eq.~(\ref{W2_coh}) is theoretically valid for incident photon energies up to about 20 TeV, above which atomic electron screening should be considered.  The reason for this number is as follows: $1/\sqrt{t}$ is the effective wavelength of a force mediator (virtual photon) that couples to a nucleus.  When this is in the order of the atomic radius, the screening effect appears.  When $k$ is large enough and the momenta of pair-produced muons are balanced, a typical size of the wavelength is $1/\sqrt{t_{\rm min}} \sim k / (4m_{\mu}^2)$, which is a typical size of an atomic radius at $k \sim $ 20~TeV.  In the current version of PHITS, the maximum transport energy of photons is 1 TeV, so we can ignore the screening effect in this implementation.

\section{Results}
\label{sec:Results}
\subsection{Comparison with the experimental data}
\label{sec:Comparison}

\begin{figure}[!ht]
\begin{center}
\includegraphics[width=15.0cm, bb=0 0 1085 710]{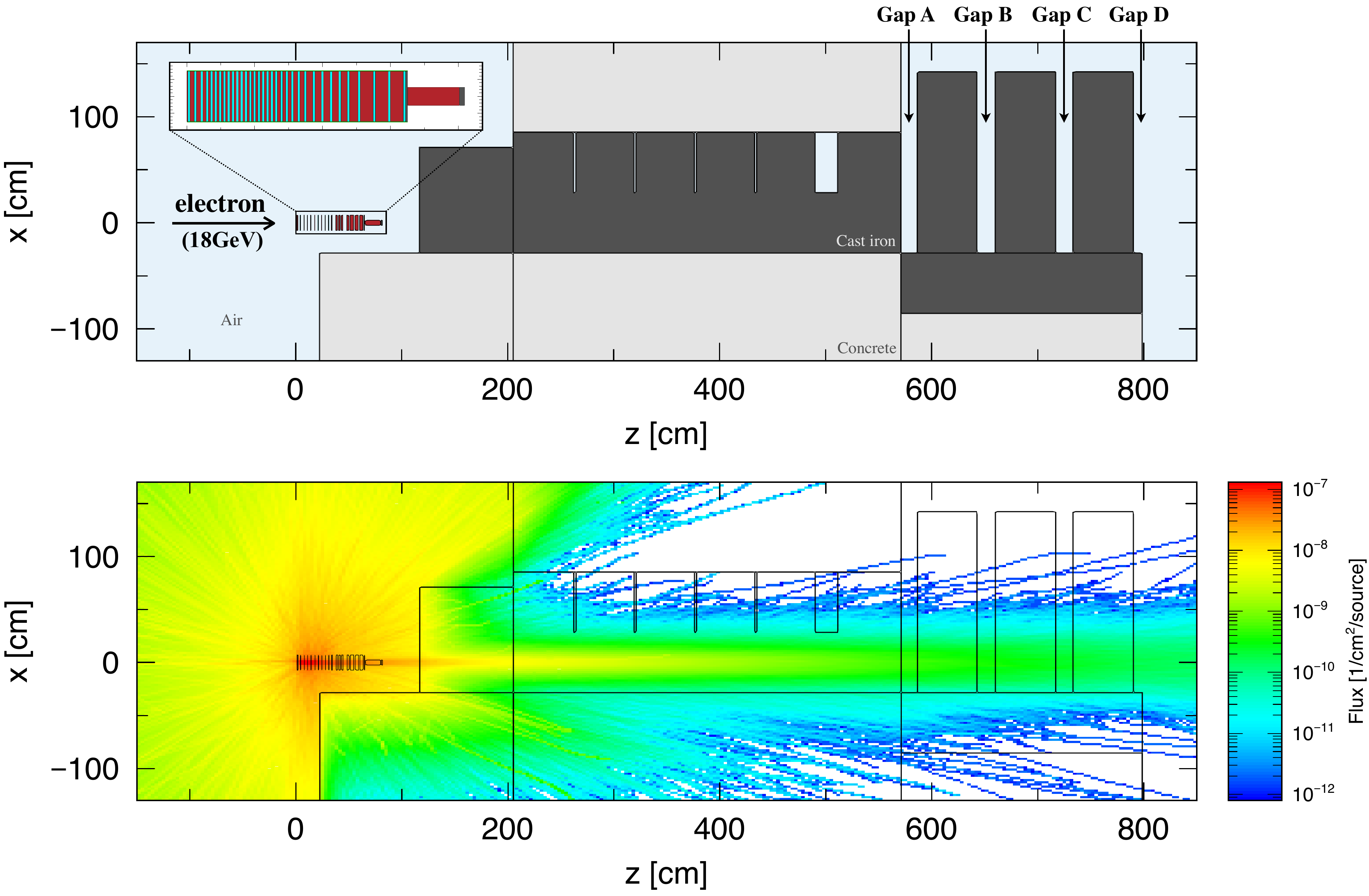}
\caption{{\footnotesize (Upper panel): Calculation geometry for the muon shielding experiment at SLAC.  
(Lower panel): A density plot of the muon flux on the geometry, calculated by {\tt PHITS}.
}}
\label{geometry}
\end{center}
\end{figure}
We verify the results of the muon pair production model implemented in {\tt PHITS} by comparing its outputs with data on the muon shielding experiment at SLAC~\cite{Nelson:1974tu, Nelson:1974tv}.  The upper panel of Figure~\ref{geometry} shows the calculation geometry of the experiment.  Electron beams with 18 GeV are injected into a beam dump, which is mainly composed of copper and water, and were muons are generated.  The muons pass through a total of 5-7 m of massive iron shielding and reach four gaps, labeled A, B, C and D.  The angular distributions of the muon fluence are measured using nuclear emulsion plates at this point.  Although not shown in the figure, the electrons pass through the air about 5 m in front of the beam dump.  Detailed geometries of the beam dump and the shieldings are described in Appendix~\ref{sec:Experimental_geometry}.

We use $6.8~\text{g/cm}^3$ as the density of the iron blocks through which muons pass.  Although the density is not mentioned in the original paper, it is important as this parameter determines the energy loss of the transported muons, thus, it affects not only the normalization of the muon fluence but also the angular distributions at the gaps.  The SLAC has been storing several the iron blocks.  We used the weights written on the blocks (18200-18400 lbs) and measured the volume directly to determine the iron block density.

The lower panel of Figure~\ref{geometry} shows the density plot of the muon flux.  Muons traveling in a forward direction are the component originating from the pair production.  In contrast, those distributed spherically around the beam dump are the ones originating from pion decays, wherein the pions are mainly created in the photo-nucleon reaction.  Despite the fact that the number of muons originating from the pion decay is about two orders of magnitude higher than that of the pair-produced one, the intense of the muon fluence in the forward direction is dominated by the pair-produced muons, even at higher energies.  This is mainly because the pion multiplicity increases with increasing incident electron or photon energy.  The range of the pair-produced muons increases roughly linearly with the incident energy, while the typical range of muons from pion decay increases weaker than linear as the incident energy is shared by multiple pions.  Also, the emission angle of the muon originating from pair production decreases linearly with incident energy, while that from the pion decay decreases less than linearly due to an increase in the invariant mass of the final state in the multiple pion production.

\begin{figure}[!ht]
\begin{center}
\includegraphics[width=8.0cm, bb=0 0 536 403]{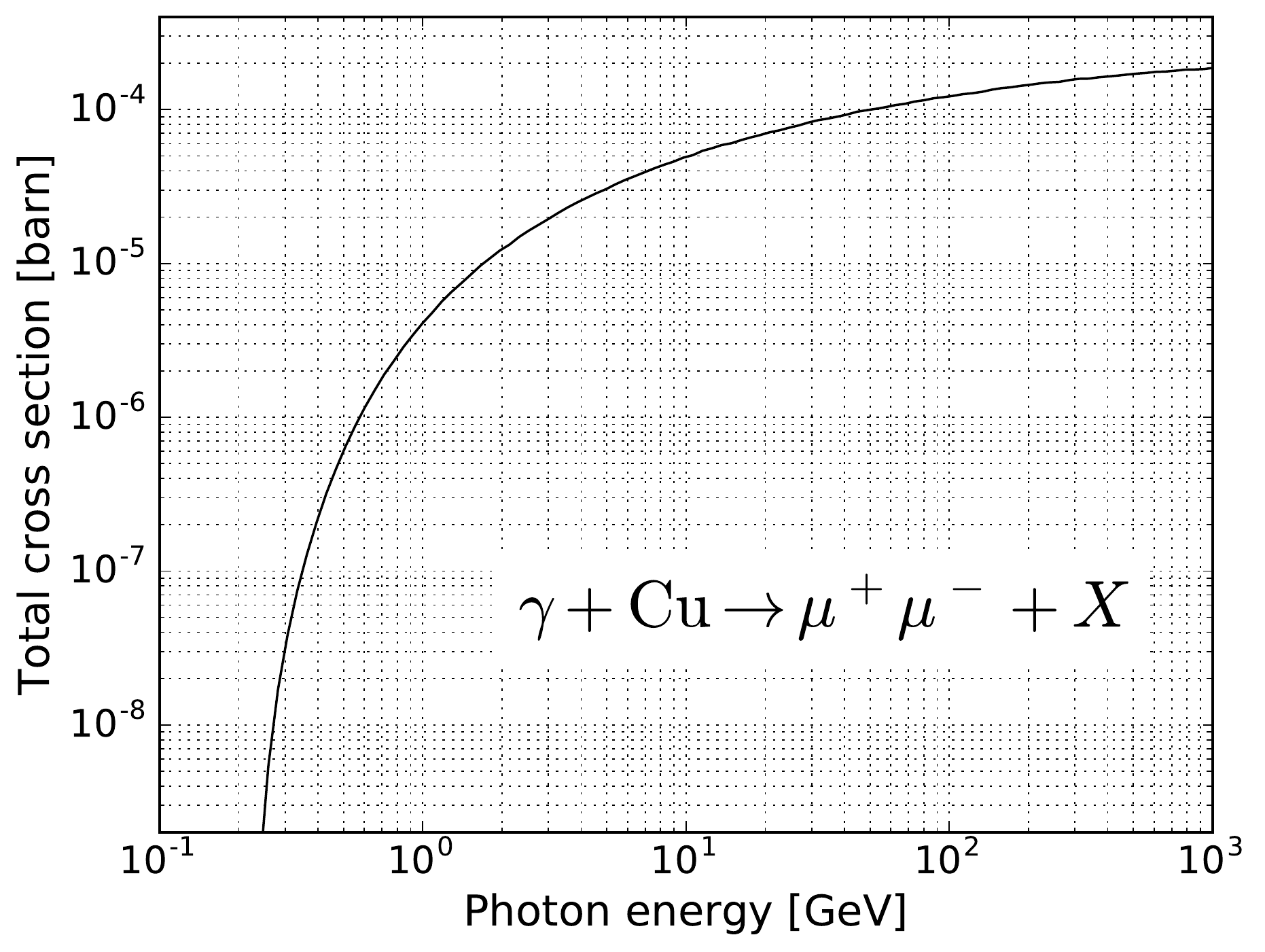}
\caption{{\footnotesize 
     An incident photon energy dependence of the total cross section for the muon pair production on a copper target.}}
\label{tot_cross}
\end{center}
\end{figure}
Before checking the numerical results of the muon fluence at the gaps, we present two supplemental numerical results.

Figure~\ref{tot_cross} shows the incident photon energy dependence of the total cross section for muon pair production on a copper target.  We observed that this cross section is about $10^{5}$ times smaller than the cross section of the electron-positron pair production.  
This suppression stems from a term of $(m_e/m_{\mu})^2$, where $m_e$ and $m_{\mu}$ are the electron and the muon masses.  This means that photons satisfying the threshold condition $k>2m_{\mu}(1+m_{\mu}/m_{\rm nucleus})$ in the electromagnetic shower generate a muon pair instead of an electron-positron pair with a probability of $\sim 10^{-5}$.  As coherent production is dominant, the total cross section is roughly proportional to $Z^2$ (see Eq.~(\ref{W2_coh})).

\begin{figure}[!ht]
\begin{center}
\includegraphics[width=8.0cm, bb=0 0 533 401]{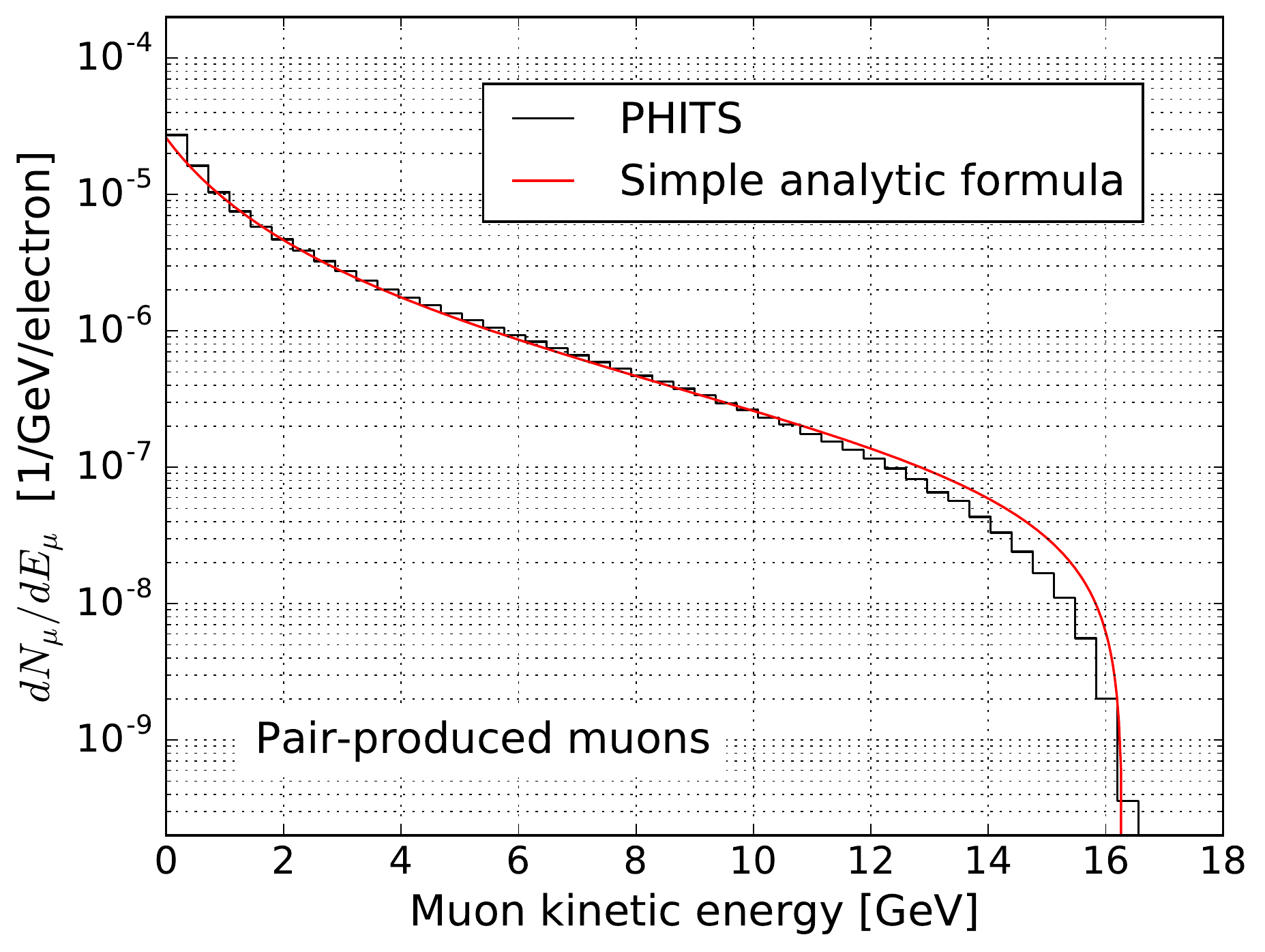}
\caption{{\footnotesize Kinetic energy distribution of the number of muons passing from the inside to the outside of the beam dump, which is normalized by the number of incident electrons.  The black curve is the result from {\tt PHITS} and the red curve is the result of a simple analytic formula given in Eq.~(\ref{eq:dNdEmu}).  Only pair-produced muons were considered.}}
\label{eng}
\end{center}
\end{figure}
Figure~\ref{eng} shows the kinetic energy distribution of the number of muons passing from the inside to the outside of the beam dump, which is normalized by the number of incident electrons.  Only pair-produced muons are considered in the figure.  The black curve is the result from {\tt PHITS} and the red curve is the one from a simple analytic formula given by 
\begin{align}
\frac{dN_{\mu}}{dE_{\mu}} \sim \frac{C_1}{\log(183 Z^{-1/3})} E_e \left( \frac{m_e}{m_{\mu}} \right)^2
    \left[ \frac{1}{(E_{\mu}+E_{\rm loss})^2} - \frac{1}{(E_e-2m_{\mu})^2} \right], \label{eq:dNdEmu}
\end{align}
where $C_1=0.572$ is Rossi's parameter of the track length of the photon~\cite{Rossi:1952kt}, $E_e$ and $E_{\mu}$ are the incident electron and muon kinetic energies, respectively, and $E_{\rm loss}$ is the energy that a muon loses in the beam dump.  Although the energy loss depends slightly on the muon energy, here, we set it universally to $E_{\rm loss} = 1.5~{\rm GeV}$, which roughly corresponds to the energy loss of muons produced at the beginning of the beam dump.  They are in good agreement, and this comparison works as a first-step verification of the {\tt PHITS} calculation.  In the derivation of the formula, we assume that when the energy of the incident photon is $k$, the energy distribution of the muon is completely flat in the region, $0<E_{\mu}<k-2m_{\mu}$.  However, phase space suppression actually occurs at the edge of the region, which is presumed to be a cause of the difference between the two curves in the high energy region.

\begin{figure}[!ht]
  \begin{center}
    \begin{tabular}{c}

      \begin{minipage}{0.5\hsize}
        \begin{center}
          \includegraphics[width=7.5cm, bb=0 0 414 402]{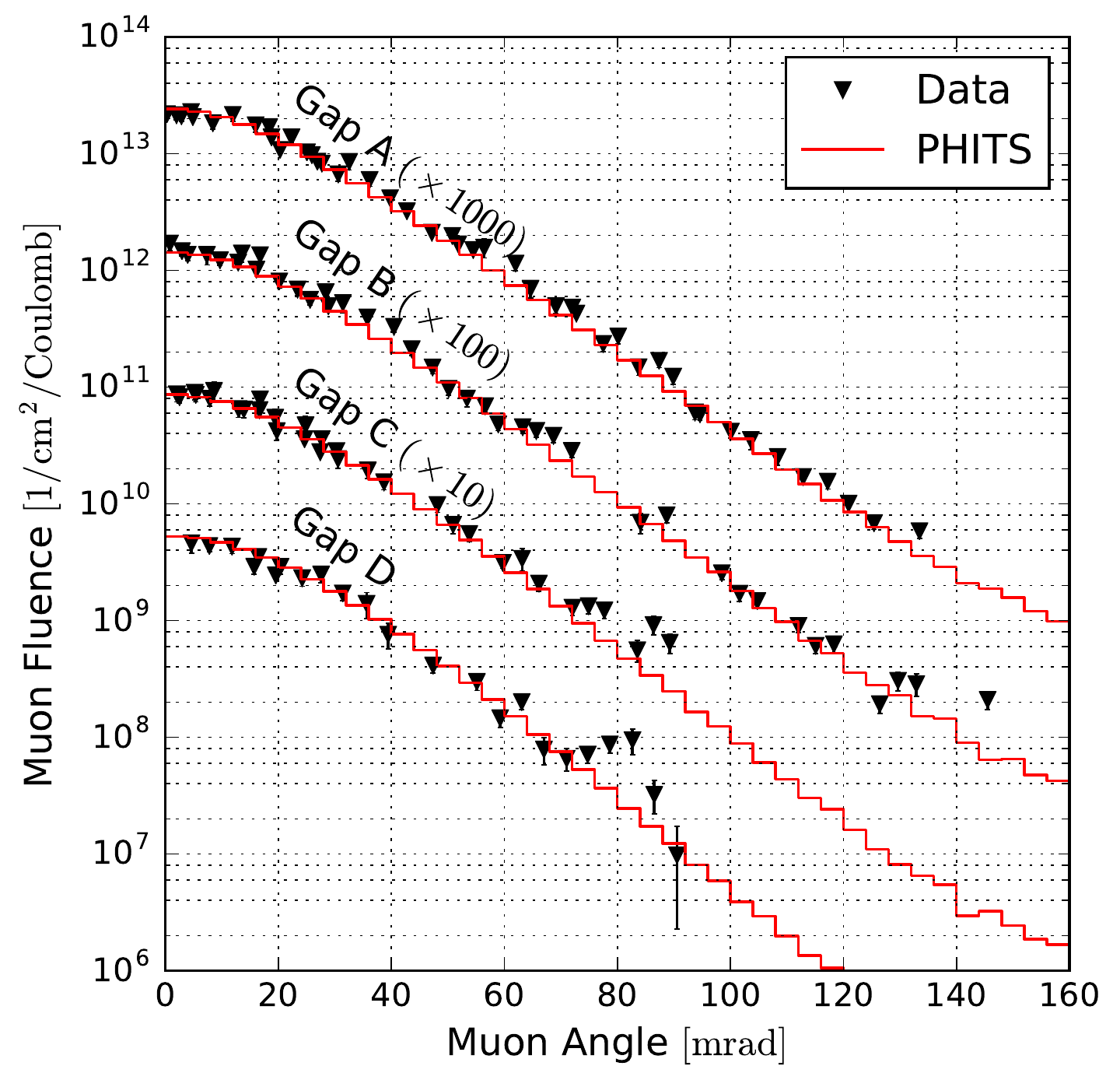}
          \hspace{1.6cm} (a)
        \end{center}
      \end{minipage}

      \begin{minipage}{0.5\hsize}
        \begin{center}
          \includegraphics[width=7.5cm, bb=0 0 430 407]{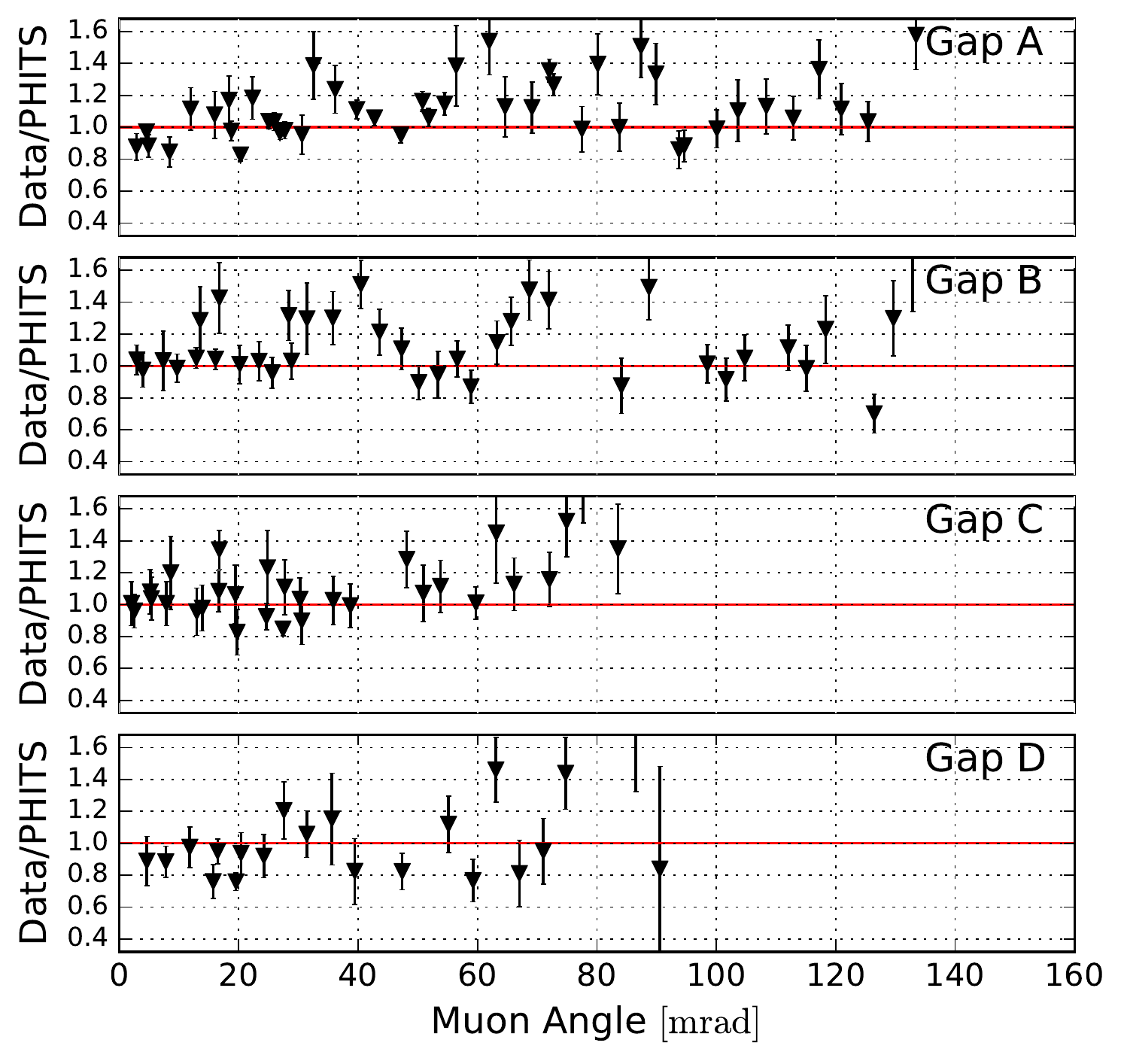}
          \hspace{1.6cm} (b)
        \end{center}
      \end{minipage}

    \end{tabular}
    \caption{(a): A comparison between the {\tt PHITS} calculations and the experimental data for the angular distributions of muon fluence at the four gaps.  The black triangle denotes the experimental data and the red histogram depicts the prediction of {\tt PHITS}. (b): The experimental data divided by the {\tt PHITS} prediction.}
    \label{fluence}
  \end{center}
\end{figure}

Figure~\ref{fluence}-(a) shows a comparison between the {\tt PHITS} calculations and the experimental data for the angular distributions of the muon fluence at the four gaps.  The black triangles show the experimental data and the red histogram shows the prediction of {\tt PHITS}.  In Figure~\ref{fluence}-(b), the experimental data is divided by the {\tt PHITS} prediction.  We found that our predictions using {\tt PHITS} code are in good agreement with the experimental data.  In some Monte-Carlo predictions, the differences to experimental data are gradually enhanced as the gap label changes from A $\to$ D~\cite{Fasso:2017cul}, however, we can see a good agreements across all gaps in our calculation.  This result is one part of the confirmation that {\tt PHITS} correctly predicts the production and transport of muons originating from electrons of order $\mathcal{O}$(10 GeV), at least.


\subsection{Fluctuations of Monte-Carlo results}
\label{sec:Discussion}

In addition to the calculation accuracy of the angular distributions of the elementary process, other factors also influence the angular distribution calculated in Sec~\ref{sec:Comparison}.  
In this subsection, we quantitatively evaluate the fluctuations of the distributions due to various factors. This may also be useful when considering the causes of known differences between Monte Carlo calculations.

We focus on the following four factors: 
\begin{enumerate}
\renewcommand{\labelenumi}{(\arabic{enumi})}
\item systematic error in multiple Coulomb scattering, 
\item differences in the approximation used in the differential cross section formula,
\item considering/not considering the incoherent productions, 
\item differences in the energy loss models. 
\end{enumerate}
In each case, we change the parameters or methods of the calculation and compare the changed calculations to the {\it base} calculation shown in Figure~\ref{fluence}.  In the base calculation, the differential cross section in the Born approximation is used, and both, coherent production and incoherent productions (quasi-elastic and inelastic scattering) are considered, and the recommended options for multiple Coulomb scattering and the energy loss model are used. 
In Table~\ref{table:parameters}, you can see the summary of the options used in the base and changed calculations.

\begin{table}[!th]
  \centering
  \caption{{\footnotesize
  Summary of calculation setup.  The base settings for $S_2$ parameter in multiple Coulomb scattering (MCS), the approximation used in the derivation of the differential cross section, the physical modes considered in the muon pair production, and the energy loss model used in the calculation are listed in the second column.  The third column shows the changed parameters or options used in Figure~\ref{fluence_mcs_ratio} and \ref{fluence_others_ratio}. 
  }}
  \vspace{5pt}
  \begin{tabular}{l|cc}
        \hline
                      &  base         &  changed  \\
        \hline
(1) $S_2$: MCS parameter  &   13.6           &  13.3, 14.4  \\
(2) Approximation           &   Born (Eq.~(\ref{eq:Born}))           &  W.W.  \\
(3) Physical mode         &   coherent+incoherent &  coherent only  \\
(4) Energy loss model   &   {\tt ATIMA}          &  {\tt SPAR}  \\
        \hline
  \end{tabular} 
  \label{table:parameters}
\end{table}

Regarding (1), multiple Coulomb scattering greatly affects the angular distribution.  
The muon fluence decreases to 1/10 of the maximum value at 50 mrad, as shown in Figure~\ref{fluence}-(a).  However, it is 20 mrad, if multiple Coulomb scattering is not considered, though this is not shown in the figure.  Therefore, correct modeling of not only the elementary process, but also of multiple Coulomb scattering is required for an accurate prediction of the distributions in the gaps.

The angular spread due to multiple Coulomb scattering is often approximately expressed in a Gaussian form.  There are several fit functions for the Gaussian RMS width, and {\tt PHITS} recommends the Lynch-Dahl formula~\cite{Lynch:1990sq} which is the one quoted by the Particle Data Group~\cite{Tanabashi:2018oca}.  The width is fitted as
\begin{align}
& \sigma = \frac{S_2~\text{MeV}}{\beta p} z \sqrt{\frac{x}{X_0}} \left( 1+\epsilon \log_{10}\frac{xz^2}{X_0 \beta^2} \right), \\
& S_2 = 13.6, \quad \epsilon = 0.088, \label{eq:LDpara}
\end{align}
where, $p$, $\beta$, and $z$ are the momentum, velocity, and charge number of the incident particle, and $x/X_0$ is the penetrating distance inside a shielding material in radiation lengths.  
It is known that the width agrees with the one in Moliere's theory describing multiple scattering with an accuracy of about $\pm 10\%$ for all materials and over a wide range of distances using parametrization of $S_2$ and $\epsilon$ in Eq.~(\ref{eq:LDpara}).  In more detail, the Lynch-Dahl width for iron targets underestimates the width by maximum 4\% compared with the width predicted in Moliere's theory for a long penetration distance, explicitly in the case of $x/X_0>1$. 
Also, when fitting a function with Gaussian-like data, it is necessary to determine the fraction of data to be used for fitting, which is called the central fraction, which we denote as $F$.  The uncertainty in $F$ leads to the error of the width.  The fitted parameters in Eq.~(\ref{eq:LDpara}) were calculated for $F = 98\%$, which corresponds to using data inbetween 2 and 3 standard deviations (SD).  Here we fix $\epsilon$ and move $F$ from 95\% (2SD) to 99.7\% (3SD) and assume the change to be a systematic error of $S_2$.  With this assumption, we estimate the error to be $S_2 = 13.6_{-0.3}^{+0.8}~(13.3<S_ 2<14.4)$.  As mentioned above, assuming Moliere's theory is correct, $S_2$ is underestimated in an iron medium, so a larger error in the positive direction is realistic.

\begin{figure}[!ht]
\begin{center}
\includegraphics[width=7.0cm, bb=0 0 420 410]{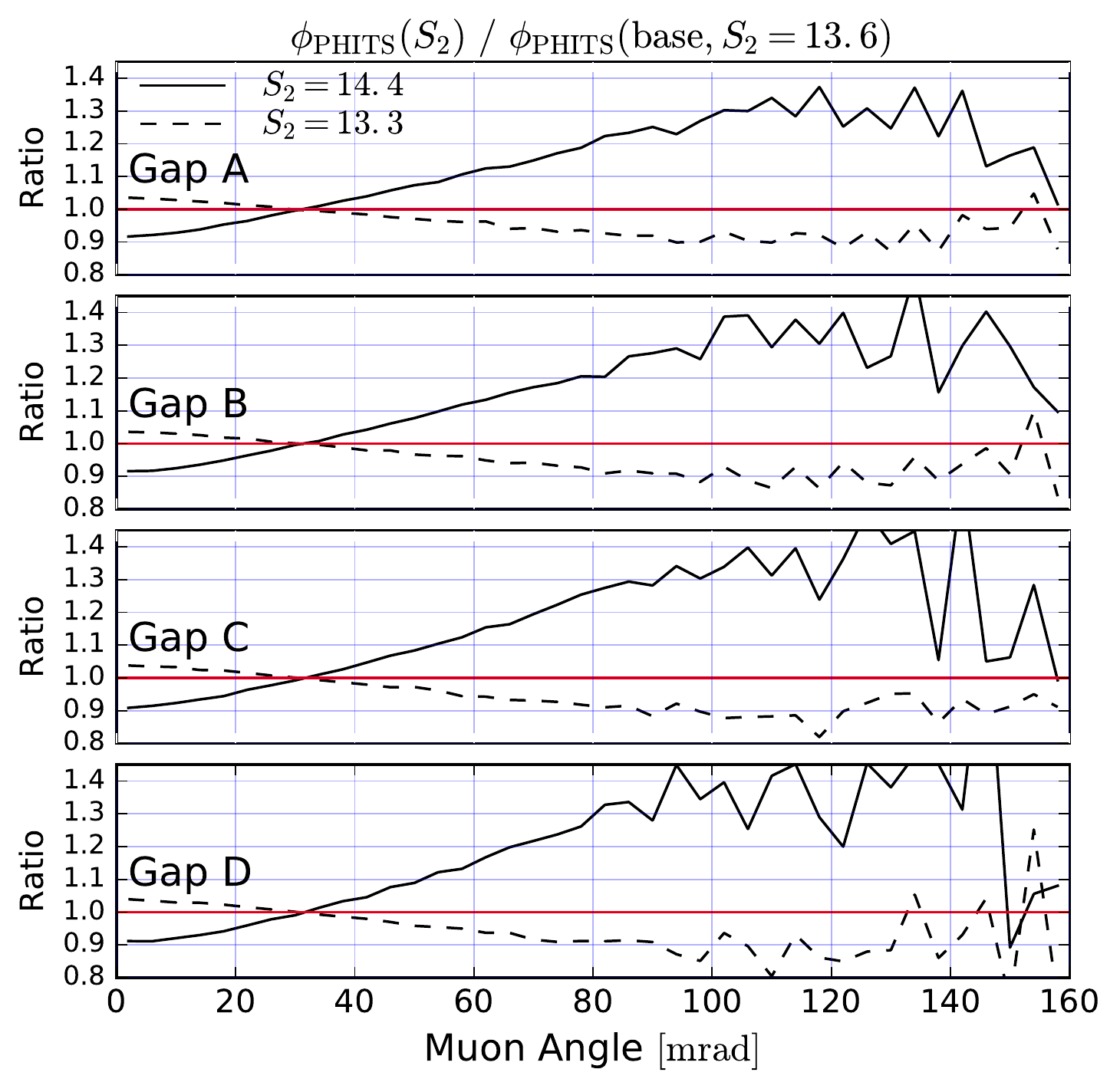}
\caption{{\footnotesize Ratio between the muon angular distributions calculated with the changed $S_2$ and the default one, where $S_2$ is set to the upper and lower values with a systematic uncertainty.}}
\label{fluence_mcs_ratio}
\end{center}
\end{figure}
We calculate the angular distributions when $S_2$ is set to the upper and lower values with the systematic uncertainty estimated in the previous paragraph.  In Figure~\ref{fluence_mcs_ratio}, the distributions divided by the base ones are shown.  
We found that the muon fluence has an uncertainty of order $\mathcal{O}(10\%)$ in a large angle region due to multiple Coulomb scattering.   For example, the uncertainty is $-10\%$ to $+30\%$ at 100 mrad.

\begin{figure}[!ht]
\begin{center}
\includegraphics[width=7.0cm, bb=0 0 420 410]{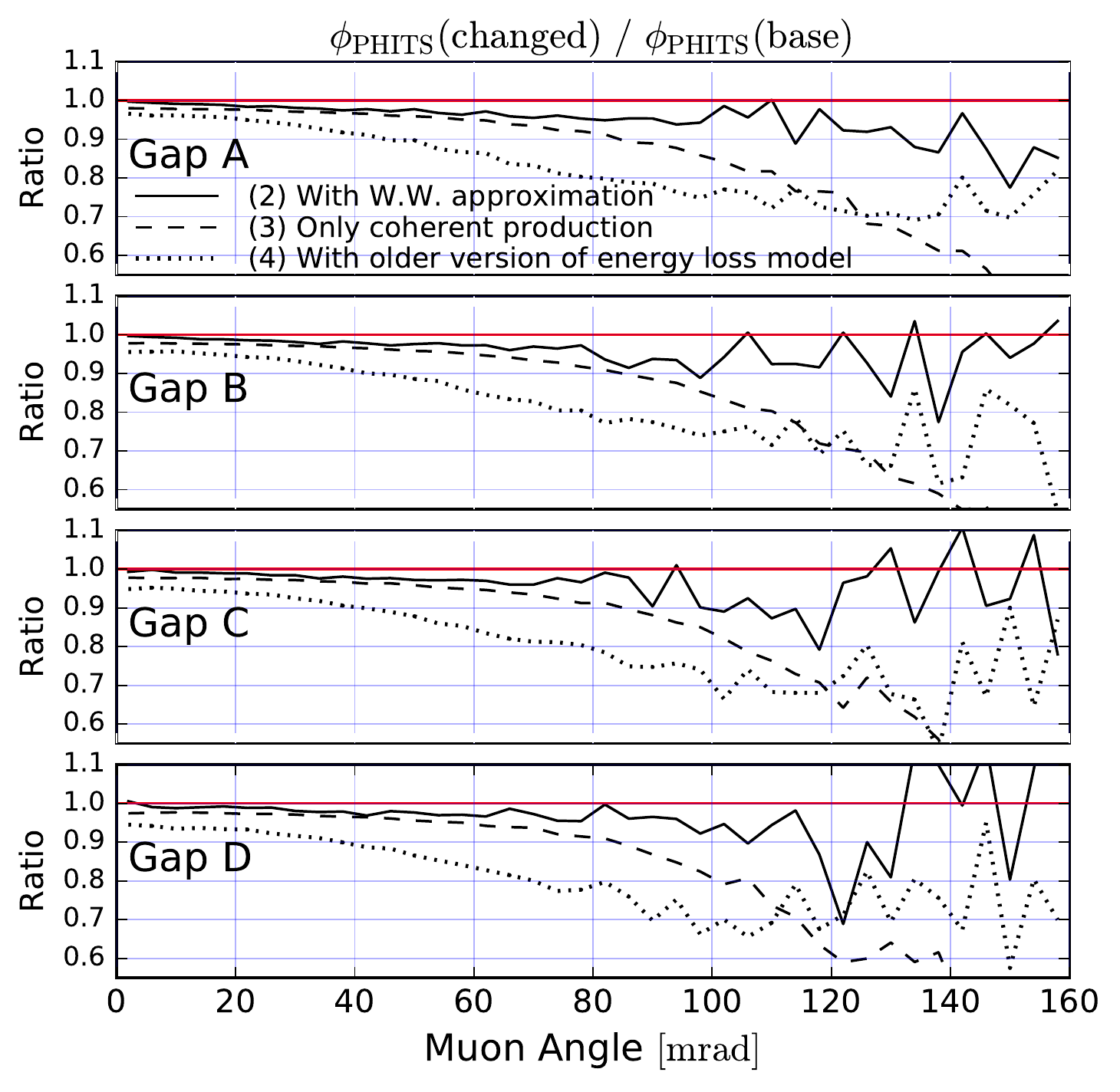}
\caption{{\footnotesize 
The solid curve is the result of using the W.W. approximation rather than the Born approximation, 
the dashed curve shows the result when the incoherent production is neglected, and 
the dotted curve is the result of using an older version of the energy loss model. 
In each case, the angular distributions at the gaps are calculated, and the ratios to the default distributions are shown.
}}
\label{fluence_others_ratio}
\end{center}
\end{figure}

Regarding (2), we introduce the Weizsacker-Williams (W.W.) approximation \cite{Kim:1973he}.  This is a simplified version of the Born approximation used in the base calculation. 
Although the Born approximation is more theoretically rigorous, the W.W. approximation has often been used in previous studies because of its simplicity.  Calculations using either approximations are compared in order to directly confirm the accuracy of the W.W. approximation.  In Figure ~\ref{fluence_others_ratio}, the solid curves show the distributions of the W.W. approximations, which are divided by those of the Born approximation.  We found that the W.W. approximation underestimated the muon fluence by a few percent when compared with the Born approximation.

Regarding (3), we confirm the effect of incoherent production on angular distributions.  In Figure~\ref{fluence_others_ratio}, the dashed curve shows the distributions in which the incoherent production is neglected, which are normalized by the base distributions.  We can see that the coherent production is dominant in the small-angle region, as mentioned in Sec.~\ref{sec:Muon_pair_production}.  Above 100 mrad, the contribution from the incoherent production gradually becomes noticeable and reaches 40\% of the total at 140 mrad.  In the original paper on the experiment~\cite{Nelson:1974tu, Nelson:1974tv}, their semi-analytic calculation of the muon fluence was 20 times smaller than the experimental data at 140 mrad for when the incoherent productions were not considered.  Our result directly confirmed that, even if the incoherent productions are considered, it is not possible to explain the discrepancy between their calculation and the experimental data, as speculated by them.

Regarding (4), we checked the existence of a typical effect coming from differences in the energy loss model. 
The recommended model used in the current version of {\tt PHITS} is a modified {\tt ATIMA}~\cite{ATIMA}.  An older model which was a default in the older {\tt PHITS} up to version {\tt 2.85} is {\tt SPAR}~\cite{Armstrong:1973ze}.  Various corrections for the stopping power have been taken into account in {\tt ATIMA}.  
In Figure~\ref{fluence_others_ratio}, the dotted curve shows the distributions from {\tt SPAR}, which are divided by those from {\tt ATIMA}.  We found that the older model predicts an about 5\% lower normalization compared to the default one, due to its slightly larger stopping power.  The difference in stopping power affects not only the normalization but also the angular spread.  As the stopping power increases, low-energy muons scattered at large angles are more likely to stop in front of the gaps.  Then, we can see that the older model predicts an about 30\% smaller muon fluence at 140 mrad compared with the base distributions.

\section{Summary}
\label{sec:Summary}

In this work, we have implemented a model of muon pair production through a real photon in {\tt PHITS}.  A differential cross section calculated in a perturbative calculation in quantum electrodynamics has been employed and three physical modes for the Bethe-Heitler process have been considered.  This work has enabled {\tt PHITS} to calculate a muon field, which is formed especially behind a target where high-energy photons, electrons, and positrons are injected.

In order to verify the muon pair production model implemented, we compared its output with data of the muon shielding experiment at SLAC.  In this experiment, data on the absolute value and angular spread of the muon flux at four depths in the muon shielding are shown.  The density $6.8~\text{g/cm}^3$ of the iron blocks through which muons pass was determined in a measurement, and used in the Monte-Carlo predictions.  We found that the predictions using {\tt PHITS} code are in good agreement with the experimental data for all depths.  This result is one part of the confirmation that {\tt PHITS} can correctly predict the production and transport of muons originating from electrons of the order $\mathcal{O}$(10 GeV) or larger.

We have also quantitatively evaluated uncertainties in the muon angular distributions due to the following: (1) systematic errors in multiple Coulomb scattering, (2) differences in the approximations used in the differential cross section formula, (3) presence/neglect of incoherent productions, or (4) differences in the energy loss models.  In this evaluation, we changed the calculation options one by one and compared all of them with the results of the base calculation of {\tt PHITS}.  
Our findings are as follows:
(1) The muon fluence has an uncertainty in the order of $\mathcal{O}(10\%)$ in the large angle region due to multiple Coulomb scattering.  Explicitly, the uncertainty is $-10\%$ to $+30\%$ at 100 mrad for the four depths. 
(2) W.W. approximation underestimates the muon fluence by a few percent compared with results using the Born approximation. 
(3) Although coherent production is dominant in the small-angle region, the contribution from incoherent productions gradually becomes noticeable above 100 mrad and reaches 40\% of the total at 140 mrad.  
(4) The differences in the energy loss models can affect the muon angular distributions.  An older model of the energy loss in {\tt PHITS} predicts an about 5\% lower normalization and a 30\% smaller muon fluence at 140 mrad compared with the base prediction of {\tt PHITS}. 
These quantitative evaluations will help in understanding the known differences in the calculations which have been determined with other Monte-Carlo codes.

\section*{Acknowledgements}
We would like to thank Walter Ralph Nelson, Shinichiro Abe, Tatsuhiko Sato, and Hirohito Yamazaki for helpful discussions and useful comments.  We also want to thank Sayed H. Rokni and other members in the Radiation Protection Department at SLAC for their hospitality during the visit at SLAC.

\appendix
\section{Experimental geometries}
\label{sec:Experimental_geometry}
In Sec.~\ref{sec:Results}, we calculated a geometry for the muon shielding experiment at SLAC.  Here we denote the details of the calculation geometry.  It was built with reference to experimental notes.  
Table~\ref{table:beamdump} shows the geometric parameters of the beam dump.  The copper and iron disks are covered with water for transport of waste heat, which is further covered with stainless steel.  The objects that make up the beam dump are represented by cylinders, and the axes of the cylinders are on the $z$-axis.  The radius, length, and $z$-coordinate minimum values for each object are listed in the table.  
Table~\ref{table:shielding} shows information on shielding designs composed of cast-iron and concrete blocks.  The objects that make up the shielding are represented by cuboids, and their faces are parallel to any of the $x$-$y$, $y$-$z$, $z$-$x$ planes.  The maximum and minimum values of the $x$, $y$ and $z$ coordinates of those objects are described in the table.  
Table~\ref{table:material} summarizes the density and composition of the materials used in the calculations.  The values in parentheses in the composition column indicate the atomic ratio of each material.

\LTcapwidth=\textwidth
\begin{longtable}{rrrrl}
  \caption{{\footnotesize
  (Unit = cm).  
  Geometry of a beam dump of the muon shielding experiment at SLAC.  
  The objects that make up the beam dump are represented by cylinders, and the axes of the cylinders are on the z-axis. The radius, length, and z-coordinate minimum values for each object are listed.
  (*1):  Copper within $0<z <64.4525~{\rm cm}$ is covered with cylindrical water.  
  (*2):Copper, water and iron within $0<z <64.4525~{\rm cm}$ are covered in steel.
  }}
  \label{table:beamdump}
  \endfirsthead
       \hline
        Material  & Radius  & $z_{\rm min}$   & Length  \\
        \hline
    Copper  &  7.4250  &   0.0000  &   0.3175  &     \\
    Copper  &  7.4250  &   0.9525  &   1.2700  &     \\
    Copper  &  7.4250  &   2.8575  &   1.2700  &     \\
    Copper  &  7.4250  &   4.7625  &   0.9525  &     \\
    Copper  &  7.4250  &   6.3500  &   0.6350  &     \\
    Copper  &  7.4250  &   7.6200  &   0.6350  &     \\
    Copper  &  7.4250  &   8.8900  &   0.6350  &     \\
    Copper  &  7.4250  &  10.1600  &   0.6350  &     \\
    Copper  &  7.4250  &  11.4300  &   0.6350  &     \\
    Copper  &  7.4250  &  12.7000  &   0.6350  &     \\
    Copper  &  7.4250  &  13.9700  &   0.6350  &     \\
    Copper  &  7.4250  &  15.2400  &   0.6350  &     \\
    Copper  &  7.4250  &  16.5100  &   0.6350  &     \\
    Copper  &  7.4250  &  17.7800  &   0.6350  &     \\
    Copper  &  7.4250  &  19.0500  &   0.6350  &     \\
    Copper  &  7.4250  &  20.3200  &   0.6350  &     \\
    Copper  &  7.4250  &  21.5900  &   0.6350  &     \\
    Copper  &  7.4250  &  22.8600  &   0.6350  &     \\
    Copper  &  7.4250  &  24.1300  &   0.6350  &     \\
    Copper  &  7.4250  &  25.4000  &   0.6350  &     \\
    Copper  &  7.4250  &  26.6700  &   0.9525  &     \\
    Copper  &  7.4250  &  28.2575  &   0.9525  &     \\
    Copper  &  7.4250  &  29.8450  &   0.9525  &     \\
    Copper  &  7.4250  &  31.4325  &   1.2700  &     \\
    Copper  &  7.4250  &  33.3375  &   1.2700  &     \\
    Copper  &  7.4250  &  35.2425  &   1.9050  &     \\
    Copper  &  7.4250  &  37.7825  &   1.9050  &     \\
    Copper  &  7.4250  &  40.3225  &   1.9050  &     \\
    Copper  &  7.4250  &  42.8625  &   1.9050  &     \\
    Copper  &  7.4250  &  45.4025  &   1.9050  &     \\
    Copper  &  7.4250  &  47.9425  &   2.5400  &     \\
    Copper  &  7.4250  &  51.1175  &   3.8100  &     \\
    Copper  &  7.4250  &  55.5625  &   3.8100  &     \\
    Copper  &  7.4250  &  60.0075  &   3.8100  &     \\
     Water  &  7.4250  &   0.0000  &  64.4525  & (*1)  \\
      Iron  &  7.4250  &  64.4525  &   0.6350  &     \\
     Steel  &  7.6200  &   0.0000  &  65.0875  & (*2)  \\
    Copper  &  2.7000  &  65.0875  &  15.2400  &     \\
      Iron  &  2.7000  &  80.3275  &   1.5875  &     \\
        \hline
\end{longtable}

\begin{table}[!h]
  \centering
  \caption{{\footnotesize
  (Unit = cm).  
  Geometry of shielding for the muon shielding experiment at SLAC. 
  The objects that make up the shielding are represented by cuboids, and their faces are parallel to any of the $x$-$y$, $y$-$z$, $z$-$x$ planes. The maximum and minimum values of the $x$, $y$ and $z$ coordinates of those objects are described. 
  (*3): There are 5 air-voids inside.
  }}
  \vspace{5pt}
  \begin{tabular}{rrrrrrrl}
  \hline
 Material  &    $x_{\rm min}$  &      $x_{\rm max}$  &     $y_{\rm min}$  &    $y_{\rm max}$  &      $z_{\rm min}$  &      $z_{\rm max}$  &       \\
        \hline
      Air  &   28.45  &   85.35   &  -200.0  &  200.0  &   262.19  &  264.255  &       \\
      Air  &   28.45  &   85.35   &  -200.0  &  200.0  &   319.09  &  321.155  &       \\
      Air  &   28.45  &   85.35   &  -200.0  &  200.0  &   375.99  &  378.055  &       \\
      Air  &   28.45  &   85.35   &  -200.0  &  200.0  &   432.89  &  434.955  &       \\
      Air  &   28.45  &   85.35   &  -106.5  &  106.5  &   489.79  &  511.380  &       \\
Cast iron  &  -28.45  &   71.125  &  -100.0  &  100.0  &   116.84  &  205.290  &       \\
Cast iron  &  -28.45  &   85.350  &  -200.0  &  200.0  &   205.29  &  571.050  &  (*3)   \\
Cast iron  &  -28.45  &  142.250  &  -100.0  &  100.0  &  586.550  &  642.430  &       \\
Cast iron  &  -28.45  &  142.250  &  -100.0  &  100.0  &  659.940  &  716.840  &       \\
Cast iron  &  -28.45  &  142.250  &  -100.0  &  100.0  &  733.205  &  790.105  &       \\
Cast iron  &  -85.35  &  -28.450  &  -100.0  &  100.0  &  571.050  &  798.650  &       \\
 Concrete  & -212.66  &   -28.45  &  -100.0  &  100.0  &    22.52  &  205.290  &       \\
 Concrete  & -212.66  &   -28.45  &  -200.0  &  200.0  &   205.29  &  571.050  &       \\
 Concrete  &   85.35  &   176.79  &  -200.0  &  200.0  &   205.29  &  571.050  &       \\
 Concrete  & -212.66  &   -85.35  &  -100.0  &  100.0  &   571.05  &  912.450  &       \\
        \hline
  \end{tabular}
  \label{table:shielding}
\end{table}

\begin{table}[!h]
  \centering
  \caption{{\footnotesize
  Density and composition of the materials used in the calculations.  
  The values in parentheses in the composition column indicate the atomic ratio of each material. 
  }}
  \vspace{5pt}
  \begin{tabular}{rrl}
        \hline
 Material  &  Density[g/cm$^3$]  &  Composition  \\
        \hline
   Copper  &   8.96           &  Cu(1.00)  \\
    Water  &   1.00           &  H(0.67), O(0.33)  \\
     Iron  &   7.87           &  Fe(1.00)  \\
    Steel  &   8.00           &  Fe(0.70), Cr(0.20), Ni(0.09), Mn(0.01)  \\
      Air  &   0.00121     &  N(0.79), O(0.21)  \\
Cast iron  &   6.8            &  Fe(0.81), C(0.14), Si(0.05)  \\
 Concrete  &   2.2            &  H(0.171), C(0.014), O(0.559), Si(0.205), \\
                  &                    &  Al(0.023), Ca(0.024), Fe(0.004) \\
        \hline
  \end{tabular} 
  \label{table:material}
\end{table}


\end{document}